\documentclass{article}

\usepackage{arxiv}

\usepackage[utf8]{inputenc} 
\usepackage[T1]{fontenc}    
\usepackage{hyperref}       
\usepackage{url}            
\usepackage{booktabs}       
\usepackage{amsfonts}       
\usepackage{nicefrac}       
\usepackage{microtype}      
\usepackage{amsmath, amssymb}
\usepackage{cleveref}       
\usepackage{lipsum}         
\usepackage{graphicx}
\usepackage{doi}
\usepackage{booktabs}
\usepackage{siunitx}
\graphicspath {{figures/}}
\usepackage[dvipsnames, table]{xcolor}

\title{Temporal stability of asymptotic suction boundary layer with spectral collocation method}

\author{ \href{https://orcid.org/0000-0001-5130-0298}{\includegraphics[scale=0.06]{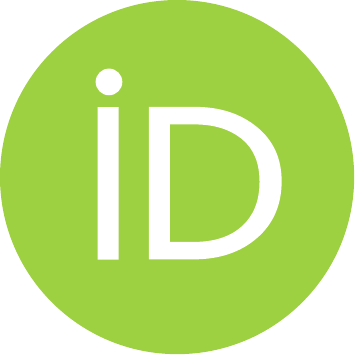}\hspace{1mm}Ressa Octavianty}\thanks{https://ressa-octavianty.gitlab.io} \\
	Department of Aviation Engineering\\
	International University Liaison Indonesia (IULI)\\
	South Tangerang, Banten, Indonesia, 15310 \\
	\texttt{ressa.octavianty@iuli.ac.id} \\
	\And
	\href{https://orcid.org/0000-0002-9009-9889}{\includegraphics[scale=0.06]{orcid.pdf}\hspace{1mm}Triwanto Simanjuntak}\thanks{https://simanjuntak-triwanto.gitlab.io/} \\
	Department of Aviation Engineering\\
	International University Liaison Indonesia (IULI)\\
	South Tangerang, Banten, Indonesia, 15310 \\
	\texttt{triwanto.simanjuntak@iuli.ac.id} \\
}

\renewcommand{\shorttitle}{Ressa Octavianty, et al.}

\hypersetup{
	pdftitle={A template for the arxiv style},
	pdfsubject={q-bio.NC, q-bio.QM},
	pdfauthor={David S.~Hippocampus, Elias D.~Striatum},
	pdfkeywords={First keyword, Second keyword, More},
}

\begin{document}
\maketitle

\begin{abstract}
	In this paper, the linear stability theory of an incompressible asymptotic suction boundary layer was studied. A small disturbance was introduced spatially in a  streamwise direction to the laminar base flow with various wavenumber $\alpha = 0.01 - 0.3$ to investigate its temporal stability. A spectral collocation method was used to solve the fourth-order ordinary differential equation (ODE) of the generalized eigenvalues problem. From the neutral stability curve, the result showed that the critical Reynolds number occurred at $Re_{c}= \num{47145}$ for $\alpha=0.161$. By taking into account that the disturbance traveled in spanwise direction, the transition can be delayed.
\end{abstract}


\section{Introduction}
\label{sec:introduction}

The asymptotic suction boundary layer is one of the canonical problems in studying laminar-to-turbulent transition. In engineering applications, finding the critical Reynolds number ($Re_{c}$) of transition flow at the boundary layer is essential to predict the aerodynamics characteristics of an object. Unlike the Blasius boundary layer, the suction boundary layer is a case that can closely describe the flow on the suction surface of an airfoil in which the $Re_{c}$ lies in the order of magnitude of $\mathcal{O}(10^{4})$.   To this date,  many theoretical, numerical, and experimental works have been done extensively to estimate the critical Reynolds number and better understand the boundary layer's transition mechanism (see the review by Reshotko  \cite{reshotko1976boundary}).

The critical Reynolds number for laminar-to-turbulent transition problems can be predicted from linear stability theory. By introducing a small three-dimensional perturbation, the Navier-Stokes equations are linearized and transformed into the fourth-order of the ordinary differential equation (ODE), known by Orr-Sommerfeld (OS) equation \cite{schlichting2016boundary, white2006viscous}. In the asymptotic suction boundary layer, Hughes and Reid \cite{hughes1965stability} solved Orr-Sommerfeld equation for various streamwise wavenumber $\alpha$ and Reynolds number $Re$ and found that the critical value was reached for $Re_{c}= \num{47047}$ with $\alpha=\num{0.163}$. Fransson and Alfredsson \cite{fransson2003disturbance} computed the modified Orr-Sommerfeld equation with a nonlinear effect for the study of suction boundary layer experiment with a porous material. The results showed that the critical Reynolds number is   $Re_{c} = \num{54382}$ for $\alpha = \num{0.1555}$. The comparison of streamwise fluctuation showed a good agreement between the theory and suction experimental work up to \SI{2}{\meter} distance from the leading edge. For a much larger distance, the results deviated significantly from the theory for the region close to the wall, $y < 8 $ times of displacement thickness ($\delta$). Later on, Tilton and Cortelezzi \cite{tilton2015stability} included the mathematical model of a porous wall to the linear stability theory and showed that the existence of the permeable structure inherently promotes a destabilizing effect. This finding verified the deviation shown in the earlier experimental work by Fransson and Alfredsson \cite{fransson2003disturbance}. Moreover, Wedin et al. \cite{wedin2015effect} showed that for the high permeability of porous wall, the critical Reynolds number was reduced to a much lower value of  $Re = \num{796}$ for $\alpha= \num{0.12}$, where the non-permeability surface gave $Re_{c} = \num{54379}$ for $\alpha = \num{0.1555}$. A recent study by Yalcin et al. \cite{yalcin2021temporal} using generalized hypergeometric functions gave a much-refined value of critical Reynolds number for linear stability of asymptotic boundary layer, that is $Re=\num{54378.62032}$.

Khater et al. \cite{khater2008chebyshev} and Malik et al. \cite{malik1985spectral}  showed that by approximating the solution using the Chebyshev polynomial as a basis and computing the derivatives at each collocation point, the ordinary system of equations (ODE) systems could be solved accurately with less computational time. Trefethen showed that the numerical computation using a function based on Chebyshev expansion is an efficient and powerful tool when dealing with a non-periodic smooth function \cite{trefethen2007computing}. The library known as chebfun is also developed extendedly to solve the linear and nonlinear periodic  ODEs \cite{wright2015extension}.   In the linear stability theory, the partial differential equation was transformed into a high-order ODEs system, i.e., Orr-Sommerfeld equation, which can be solved using the spectral collocation methods with chebfun library implementation. In this research, we explore the application of this method to solve the incompressible asymptotic suction boundary layer with two-dimensional disturbances to determine the critical Reynolds number, the disturbance growth, and the corresponding unstable modes of the temporal stability.

	\section{Numerical Method}
	\label{sec:numerical-method}

\begin{figure}[h]
		\centerline{\includegraphics[scale=0.35]{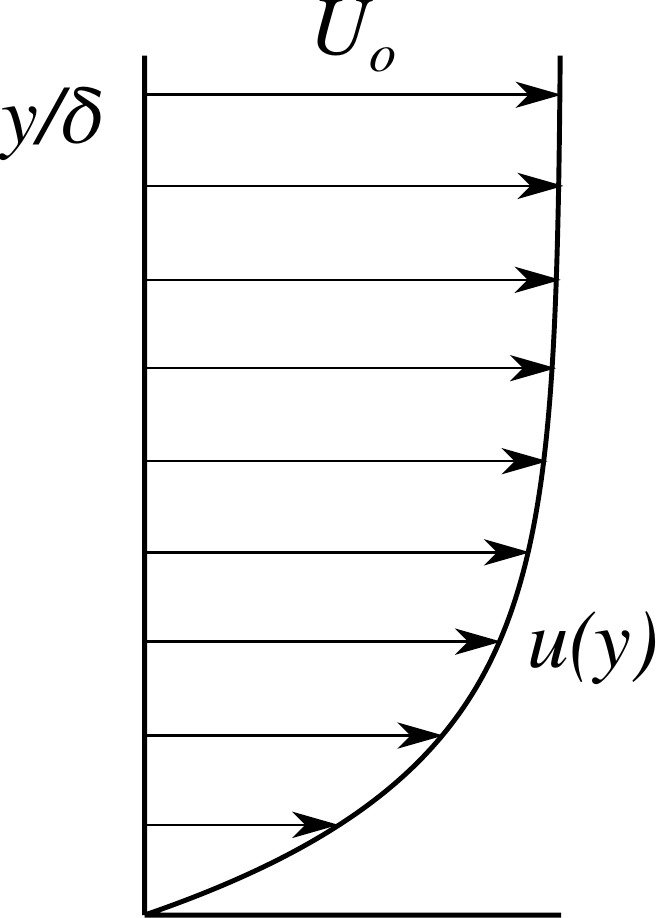}}
		\caption{The base flow of asymptotic suction boundary layer.}
		\label{figure-bl}
	\end{figure}

	In this analysis, the base flow of asymptotic boundary layer shown in Fig.~\ref{figure-bl} can be simply written as:
	\begin{equation}
		\label{eq:1}
		u(y) = U_{0}(1-e^{-yu_{s}/\nu})
	\end{equation}
	where $u$, $U_{0}$, $u_{s}$, and $\nu$ denote the streamwise velocity within the boundary layer, freestream velocity, suction velocity, and dynamic viscosity, respectively.
	Throughout the study, it was assumed that the viscous force is equal with the suction force to enable the delay of boundary layer growth. The Reynolds number
	based on the displacement thickness $\delta$ is defined as $Re = \dfrac{U_{0} \delta} {\nu}$, where $\delta = \dfrac{\nu}{u_{s}}$. The Reynolds number can be simply written as $Re = \dfrac{U_{0}}{u_{s}}$.
	A small disturbance in the form of a traveling wave in three-dimensional directions defined by $\varphi (x, y, z, t) = \psi(y)~ e^{i (\alpha x + \beta z - c t)}$ is introduced to the Navier Stokes equations. Note that $\psi$ is the stream function of the disturbances, where $\alpha$ and $\beta$ denote the spatial wavenumber of disturbance in streamwise and spanwise directions, respectively, and $c$ is the phase velocity of the disturbance.

	After linearizing and assuming that the rate of change of transverse velocity fluctuation $v^{\prime}$ in streamwise direction is very small, such that  $\frac{\partial v^{\prime}}{\partial x} \lll \frac{\partial u^{\prime}}{\partial y}$,
	the Navier Stokes equations can be transformed into fourth-order ordinary differential equation (ODE). This form is famously known as Orr-Sommerfeld equation and can be written in a non-dimensional one as follows:
	\begin{equation}
		\label{eq:2}
		\left(\bar{U} - \bar{c}\right)\left(\bar{\psi}^{\prime\prime}
		- \alpha^2\bar{\psi}\right) - \bar{U}^{\prime\prime}\bar{\psi}
		+ \frac{i}{\alpha Re}\left(\bar{\psi}^{\prime\prime\prime\prime}
		- 2\alpha^2\bar{\psi}^{\prime\prime} + \alpha^2\bar{\psi}\right)
		= 0
	\end{equation}
	where the prime notation represents the variable derivative to $y$.

	To solve Eq.~\ref{eq:2}, particular boundary conditions were applied, that is
	$\psi = \psi^{\prime}=0 $ at the wall ($y=0$) and
	$\psi = \psi^{\prime} \rightarrow 0 $ at $y \rightarrow+\infty$. Using
	differential operator $\mathcal{D} = \partial/\partial y$, Eq.~\ref{eq:2} can be
	written as:
	\begin{equation} \label{eq:4a}
		\left(\bar{U} - \bar{c}\right) \left(\mathcal{D}^{2} - k^{2}\right) \bar\psi - \bar{U}^{\prime\prime}\bar{\psi}
		+ \frac{i}{\alpha Re}\left( \mathcal{D}^{2} - k^{2}\right)^{2} \bar \psi = 0
	\end{equation}
	which then can be rearranged into,
	\begin{equation}
		\label{eq:4b}
		(\mathcal{D}^{2}- k ^{2}) \left[ -\frac{1}{Re} (\mathcal{D}^{2}-k^{2}) + i \alpha \bar{U} \right] \bar \psi - i \alpha \bar{U}^{\prime\prime} \bar \psi  = i \omega \left(\mathcal{D}^{2}- k ^{2}\right) \bar \psi
	\end{equation}

	The $k$ represents the relation between $\alpha$ and $\beta$, where $k^{2} = \alpha^{2} + \beta^{2}$. For two-dimensional disturbances, $k$ is proportional to $\alpha$.  In three dimension, the Squire transformation was needed to accompany Orr-Sommerfeld equation to represent the vorticity in $x-z$ plane which is written as:
	\begin{equation}
		\label{eq:7}
		\left( - \frac{1}{Re} (\mathcal{D}^{2} - k^{2}) +  i \alpha \bar{U} \right) \bar \eta + i \beta \bar{U}^{\prime} \bar \psi  = i \omega \bar \eta
	\end{equation}
	Equations~\ref{eq:4b} and ~\ref{eq:7} can be written in generalized eigenvalue problem according to a linear system of equations (see Schmid and Henningson for
	detail \cite{schmid2002stability}):
	\begin{equation}
		\label{eq:5}
		\mathcal{A} ~ \boldsymbol{\bar \Psi} = \lambda ~\mathcal{B}~\boldsymbol{\bar \Psi}
	\end{equation}
	Here,
		\begin{alignat}{2}
			\label{eq:8}
			\mathcal{A}=
			\begin{pmatrix}
				A_{11}   & 0      \\
				A_{{21}} & A_{22}
			\end{pmatrix},
			 & \qquad
			\mathcal{B}=
			\begin{pmatrix}
				B_{11} & 0      \\
				0      & B_{22}
			\end{pmatrix},
		\end{alignat}
	and where $\lambda$, and  $\boldsymbol{\Psi}=\begin{pmatrix}
			\bar \psi \\
			\bar \eta
		\end{pmatrix}$
	represents eigenvalues, and eigenvectors respectively. The matrix component of $\mathcal{A}$ and $\mathcal{B}$  can be defined as:
		\begin{subequations}
			\begin{align}
				A_{11} & = (\mathcal{D}^{2}- k ^{2}) \left(- \frac{1}{Re} (\mathcal{D}^{2}- k^{2}) + i \alpha \bar{U} \right) - i \alpha \bar{U}^{\prime \prime} \\
				A_{21} & = i \beta \bar{U}^{\prime}                                                                                                              \\
				A_{22} & =  - \frac{1}{Re} (\mathcal{D}^{2}- k^{2}) + i \alpha \bar{U}                                                                           \\
				B_{11} & = \mathcal{D}^{2}- k ^{2}                                                                                                               \\
				B_{22} & = I
			\end{align}
		\end{subequations}
	The eigenvalues $\lambda$ is defined by $\lambda= i \omega$ where $\omega$ is the frequency as a function of spatial wavenumber $\alpha$ and phase velocity $c$, that is $\omega = \alpha c$. In this study,  temporal stability analysis of asymptotic suction flow was discussed by assuming $\alpha$ as a positive real number and $c$ is a complex number, $c = c_{r} + ic_{i}$. The temporal growth of disturbance will amplify when the $c_{i}$ becomes positive, indicating the onset of instability for the corresponding flow.
	\begin{figure}[h!]
		\begin{minipage}[t]{.48\textwidth}
			\centering
			\includegraphics[width=1.05\textwidth]{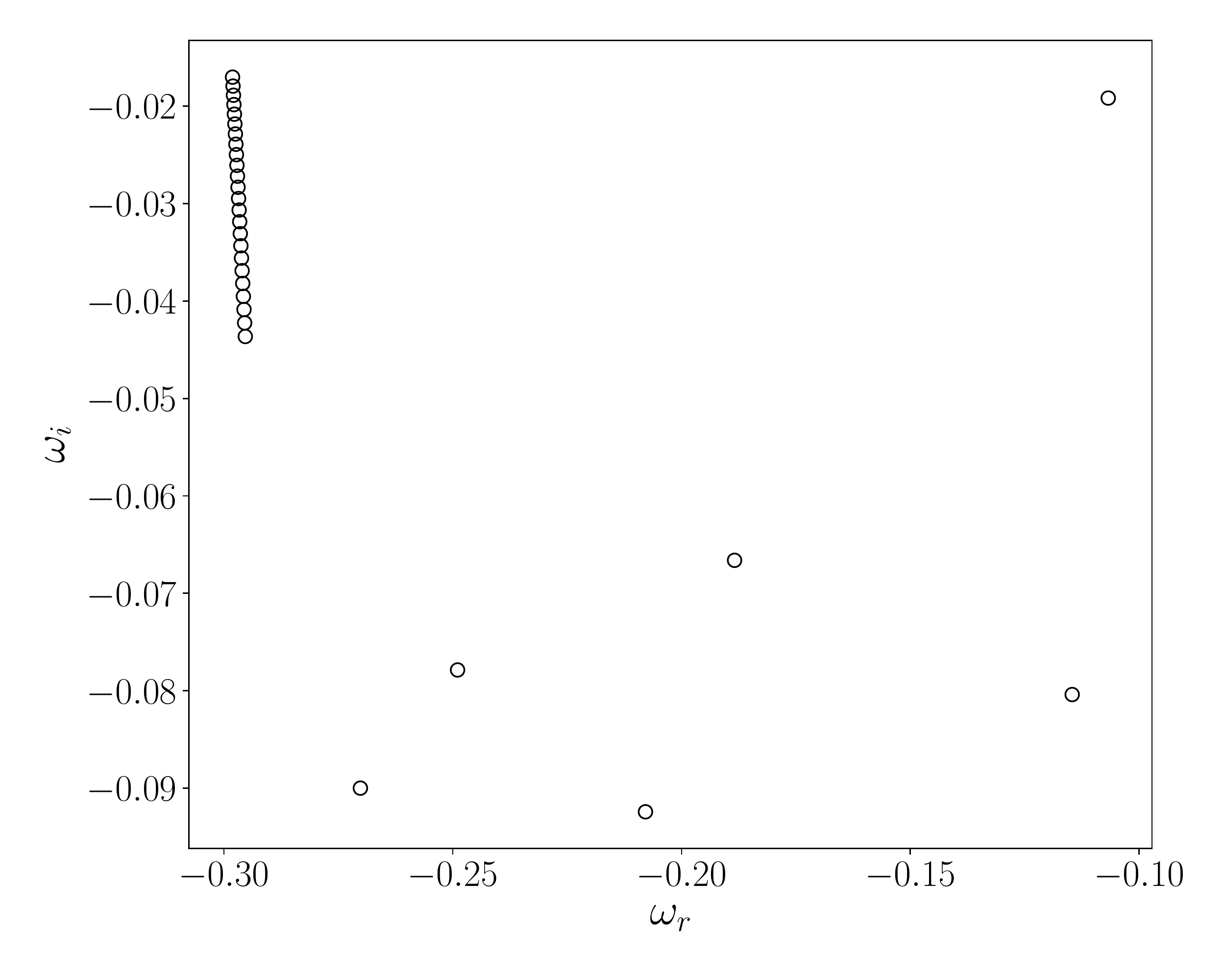}
		\end{minipage}
		\hfill
		\begin{minipage}[t]{.48\textwidth}
			\centering
			\includegraphics[width=1.05\textwidth]{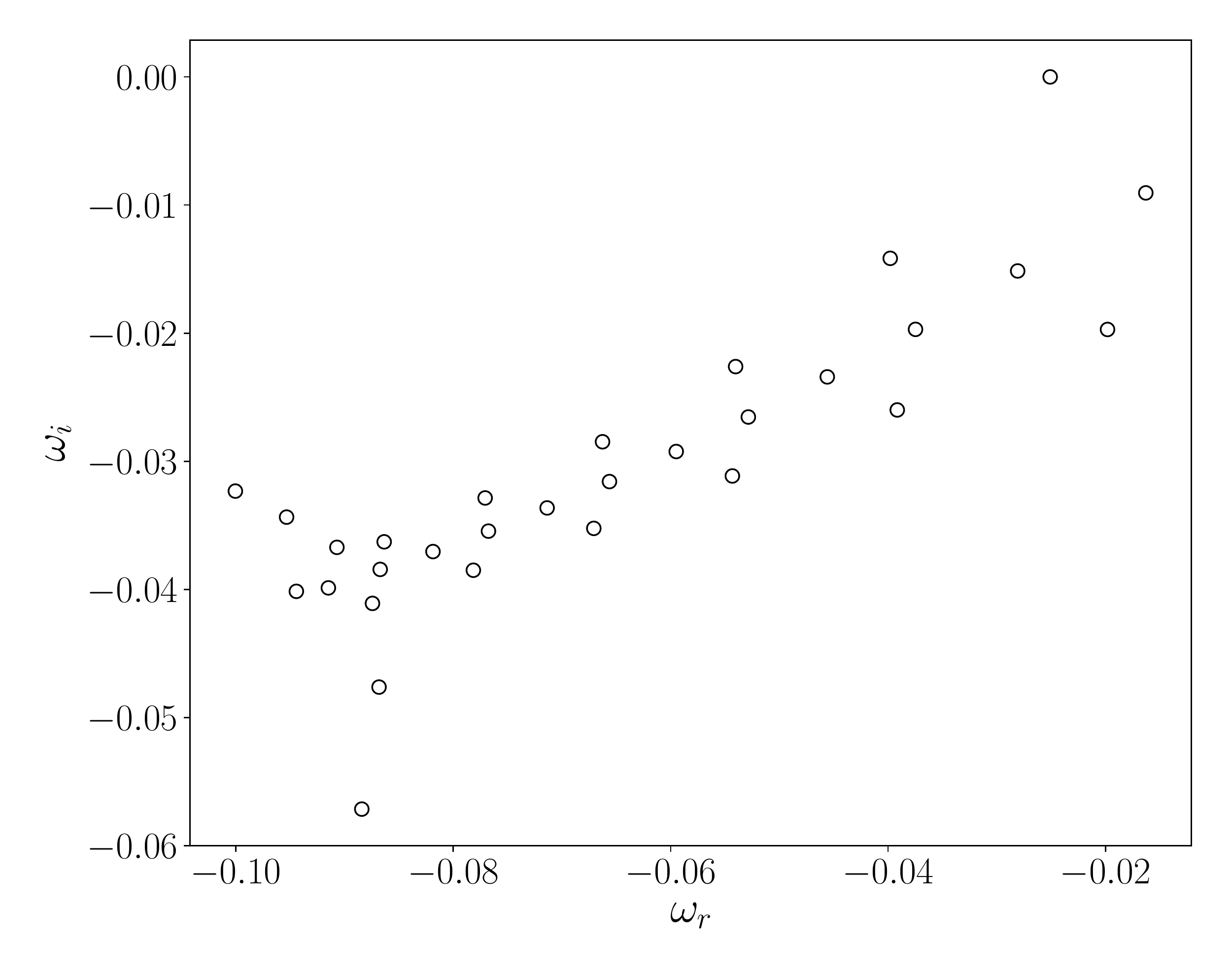}
		\end{minipage}
		\caption{Complex frequency of asymptotic suction boundary layer for $Re = \num{1000}$ at $\alpha=0.3$
			(left figure) and $Re=\num{47145}$ at $\alpha=0.161$ (right figure).}
		\label{figurethree}
	\end{figure}

	Numerically, the generalized eigenvalues problem in Eq.~\ref{eq:5} was solved using the spectral collocation
	method by using chebfun library available in MATLAB. In this library, the
	Chebyshev polynomials expansion were used to approximate the solution
	$\bar{\psi}$ and $\bar \eta$ at each collocation point, such that:
	\begin{equation}
		\label{eq:3}
		\left(\bar  \psi, \bar{\eta}\right)~ (y_{i}) = \sum_{n=0}^{N}a_{n}T_{n}(y_{i})
	\end{equation}
	where $a_n$ and $T_{n}$ are the constants and polynomial basis of Chebychev expansion. By default,  $y $ is $-1 \leq y \leq 1$  in Chebyshev domain and is
	defined by $y_{i} = - \cos(i \pi / N)$ for $0 \leq i \leq N$. In this computation, thirty eigenvalues and corresponding eigenvectors were calculated in the domain  $ y = 0 \leq y \leq 30$  which represent the normal modes of the stream function and normal vorticity. All the computation were carried out for $\alpha = 0.01 - 0.3$  and $\beta=0$ for \num{1225}  Reynolds numbers values ranging at $Re = 10^{3}-10^{8}$. Additional evaluation of eigenvalues at critical Reynolds number with $\beta= 0.05$ are also analyzed to see the effect of spanwise disturbance to the stability of suction boundary layer.

	\section{Results and Discussions}
	\label{sec:results-discussions}

	\begin{figure}[ht!]
		\centerline{\includegraphics[scale=0.55]{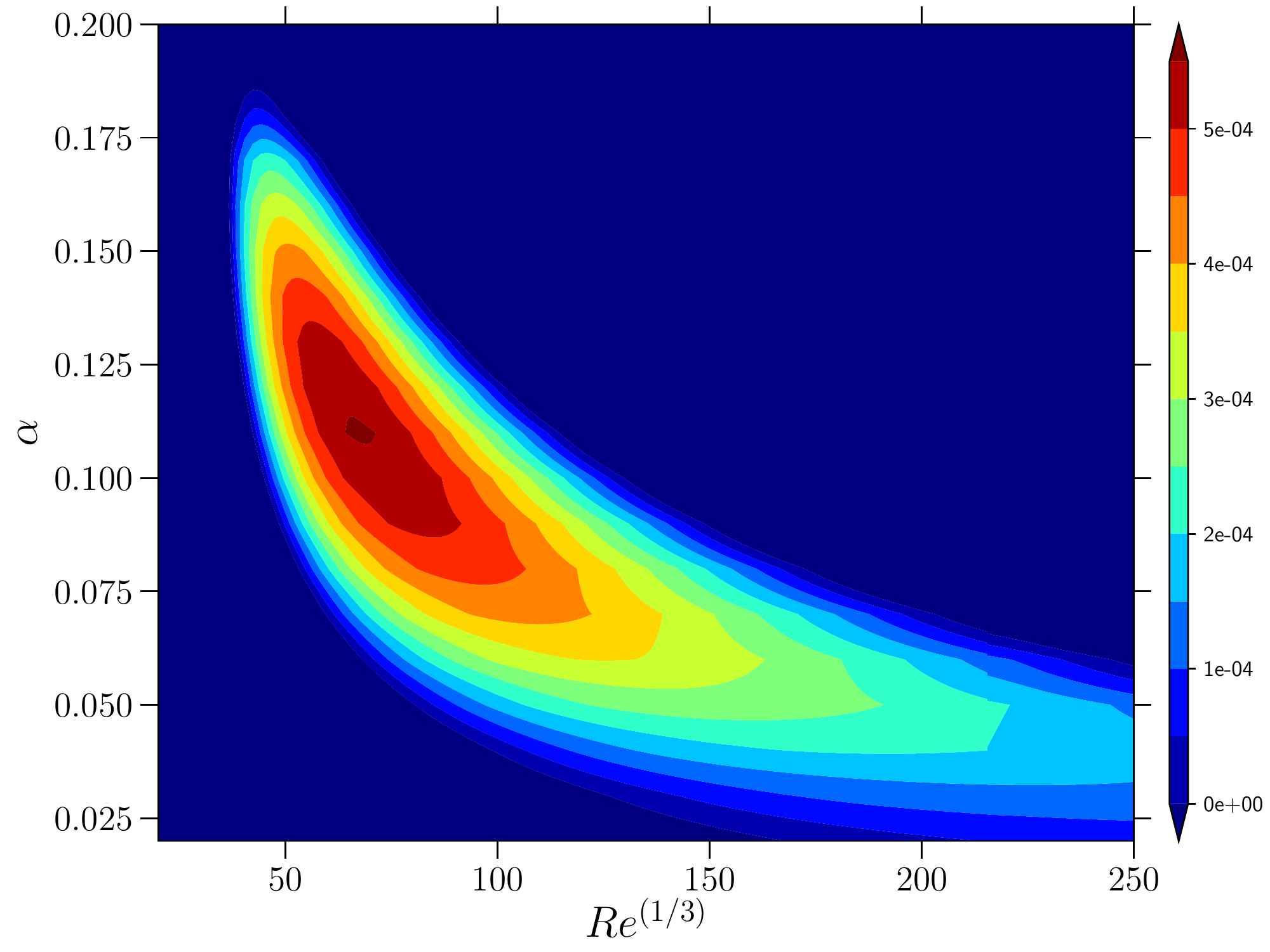}}
		\caption{Temporal stability of two-dimensional asymptotic suction boundary layer
			flow for various $\alpha$ and $Re$. The contour values represent the
			imaginary part of frequency $\omega_{i}$.}
		\label{figureone}
	\end{figure}

	Figure~\ref{figurethree} shows the eigenvalues of Orr-Sommerfeld equation with two-dimensional disturbances for $Re= \num{1000}$ and $\num{47145}$ at
	$\alpha = 0.3$ and $0.161$, respectively. From this complex eigenvalues number, information regarding phase velocity can also be obtained. At $Re=\num{47145}$ and $\alpha=0.161$, one of the modes has a  positive imaginary component of eigenvalues $\omega_{i}$  which indicates the
	onset of instability in the suction asymptotic boundary layer flow. Correspondingly, at this condition,  the critical phase velocity showed by the real number of $c$  is about  $\num{0.15588}$, which agrees with those values calculated by Hughes and Reid~\cite{hughes1965stability}. For another condition at $Re=\num{1000}$ and $\alpha=0.3$, the result shows that all stability modes of disturbances waves are dampened temporally.

	\begin{figure}[ht!]
		\centerline{\includegraphics[scale=0.5]{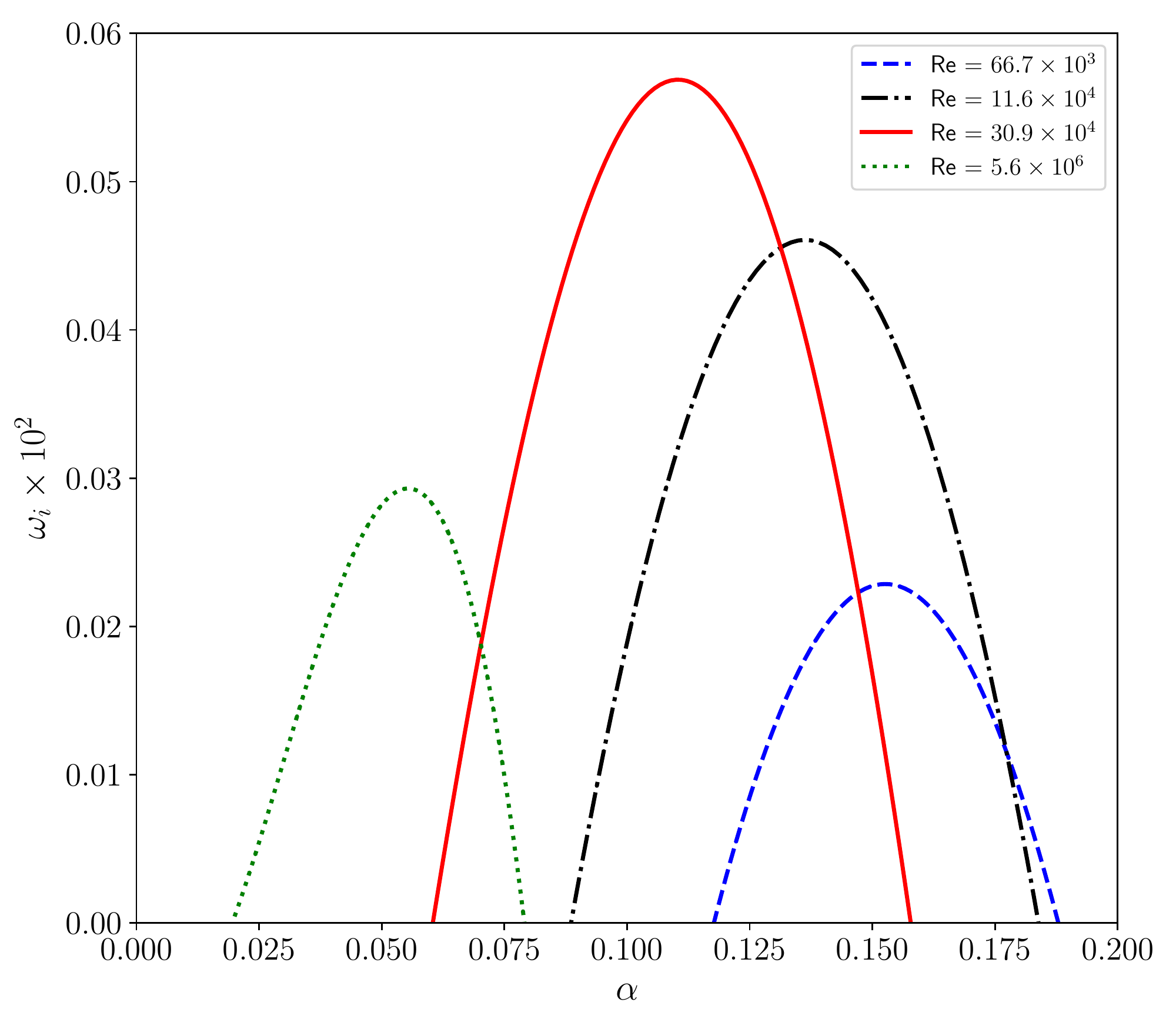}}
		\caption{Temporal growth rate of asymptotic suction boundary layer for various Reynolds numbers, $Re$.}
		\label{figuretwo}
	\end{figure}

	Figure~\ref{figureone} represents the temporal stability contour of asymptotic
	suction boundary layer flow at various Reynolds number $Re$. The contour values
	represent the imaginary components of the phase velocity, $c_{i}$. In this
	figure, the neutral curve  bounded by the value $c_{i} = 0$. From the neutral curve, it is shown that the critical Reynolds number occurred at $Re\sim \num{47145}$ for $\alpha \sim 0.161$. Comparing to previous study
	by Hughes and Reid~\cite{hughes1965stability}, the results is slightly lower
	where the Reynolds number critical for the instability occur at $Re \sim \num{47047}$
	at $\alpha=0.163$.

	The growth rate of a disturbance at constant Reynolds number $Re$ is shown in Fig.~\ref{figuretwo}. Here, a typical Reynolds number was selected to represent a condition where instability occurs in the suction boundary layer. The amplification factor reached maximum for $Re \sim 3.1 \times 10^{5}$. At this condition, the streamwise wavenumber  $\alpha < 0.06$ and  $\alpha > 0.155$ will be dampened.

	\begin{figure}[h!]
		\begin{minipage}[t]{.48\textwidth}
			\centering
			\includegraphics[width=1.\textwidth]{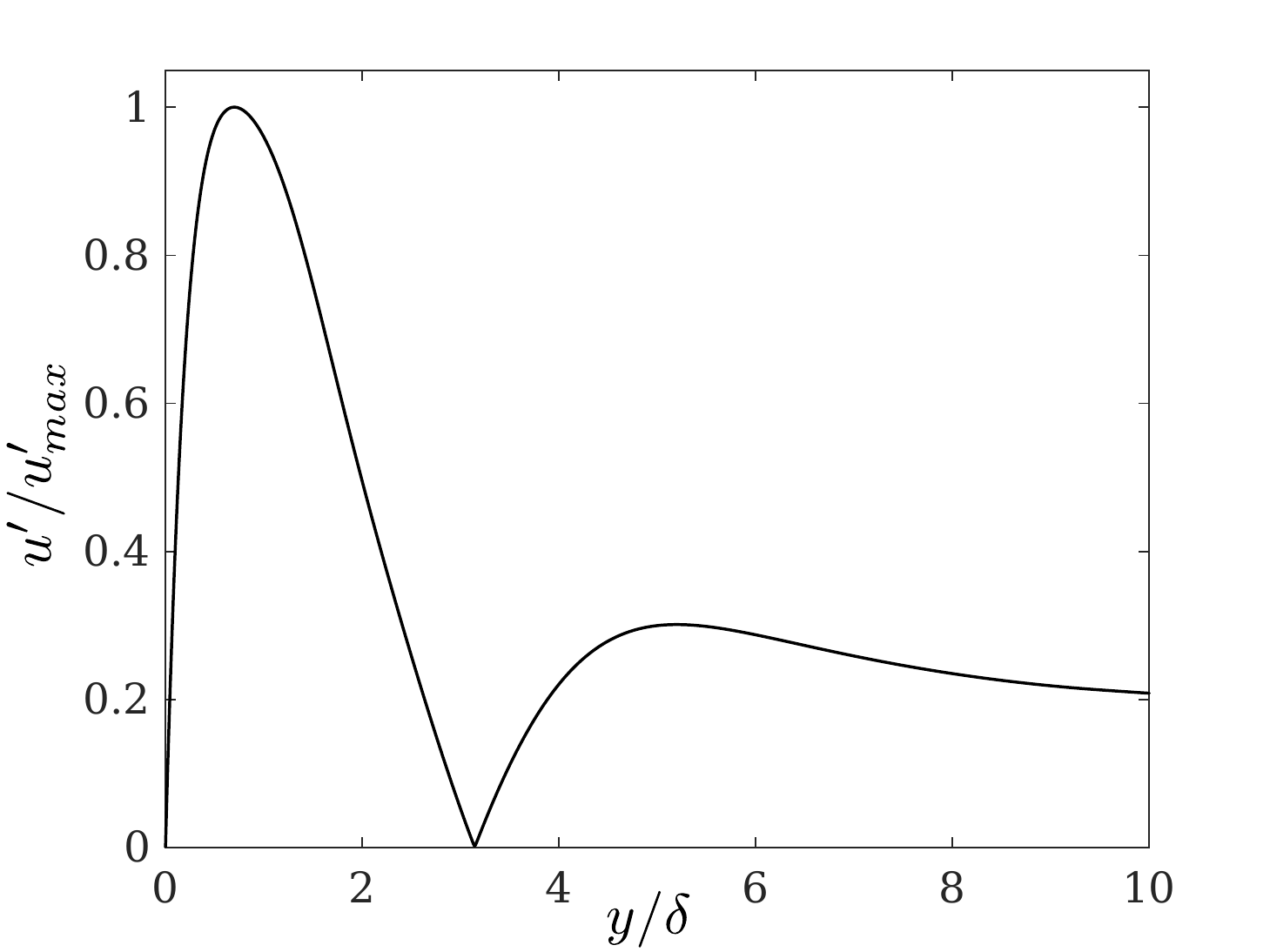}
		\end{minipage}
		\hfill
		\begin{minipage}[t]{.48\textwidth}
			\centering
			\includegraphics[width=1.\textwidth]{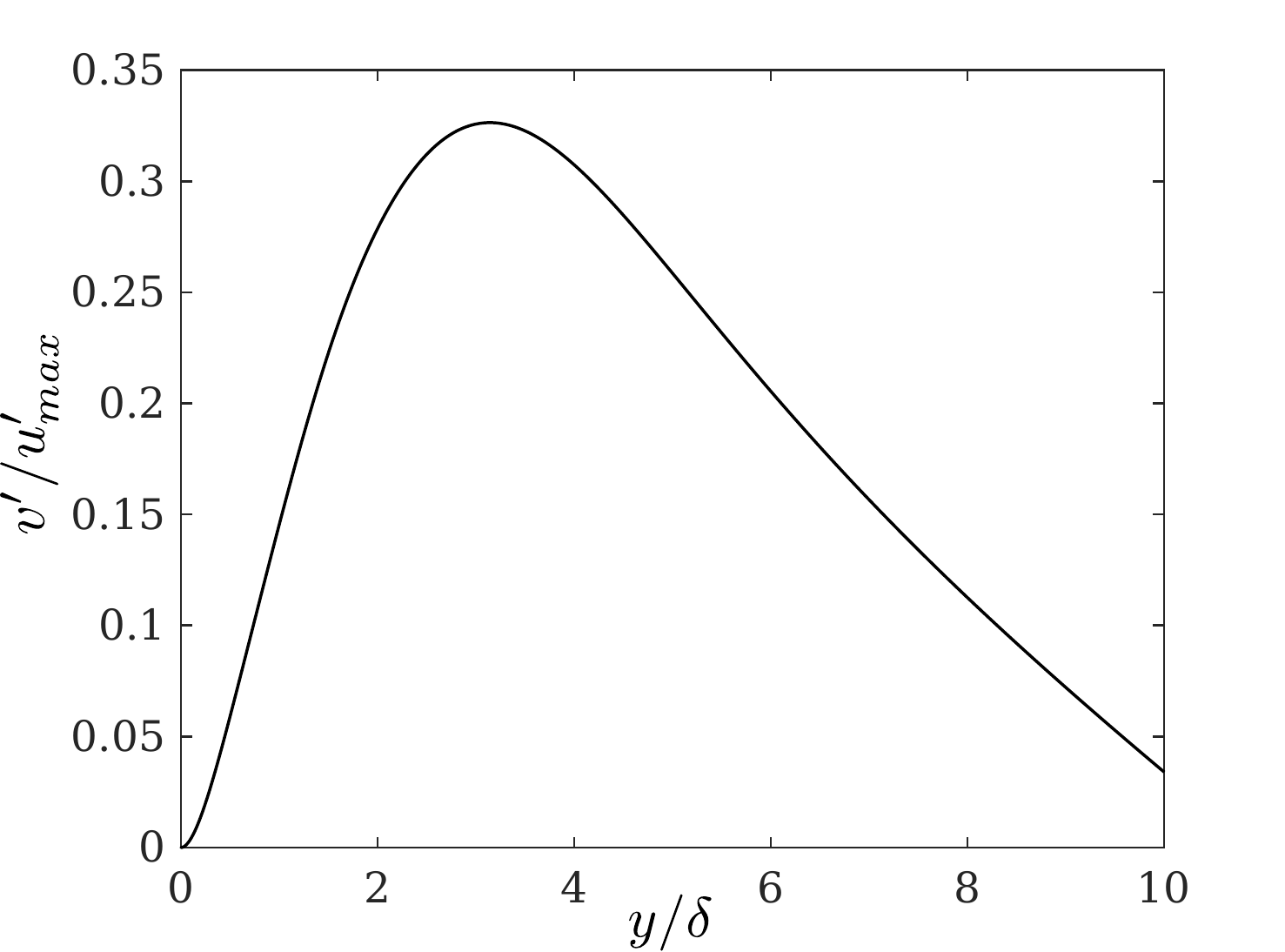}
		\end{minipage}
		\caption{The streamwise  $u^{\prime}$  (left figure) and  tranverse velocity
			fluctuations $v^{\prime}$ (right figure) for Blasius boundary layer at $Re_{\delta}=580$ and $\alpha=0.179$.}
		\label{figurefive}
	\end{figure}

	To evaluate the eigenfunctions using the spectral collocation method, validation of Blasius boundary layer was also calculated for  $Re=\num{580}$ and $\alpha=0.179$ as shown in Figure \ref{figurefive}. At this condition, the unstable mode was amplified; thus, instability occurred.  The results of streamwise $u^{\prime}$ and transverse velocity fluctuations $v^{\prime}$  are the same as those results using compound matrix solved with Runge-Kutta method \cite{ng1979initial,criminale2018theory}.

	\begin{figure}[ht!]
		\centerline{\includegraphics[scale=0.5]{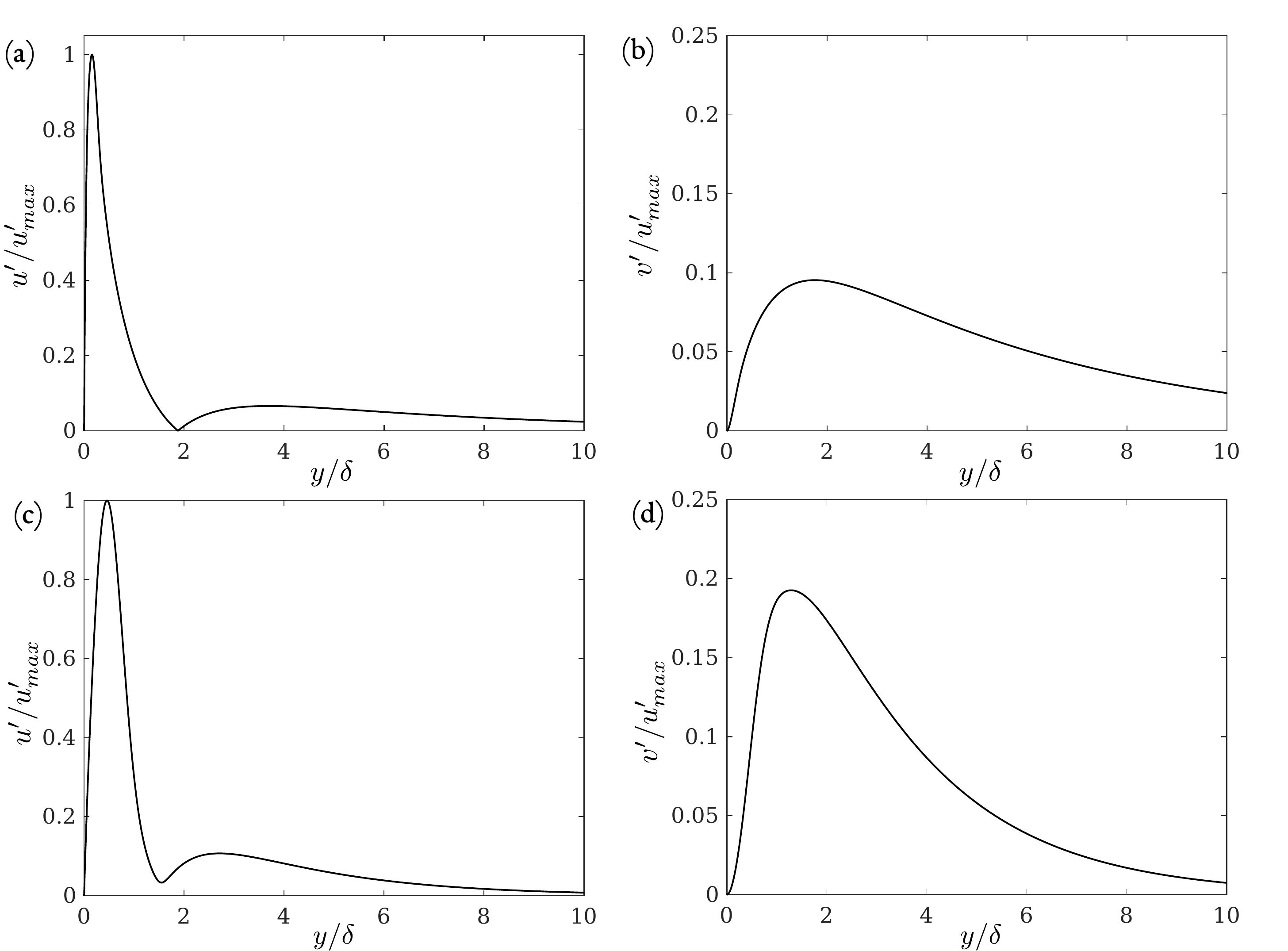}}
		\caption{The streamwise  $u^{\prime}$  and  tranverse velocity fluctuations
			$v^{\prime}$ at  critical value of $Re=\num {47145}$ and $\alpha=0.161$ in
			(a) and (b) and at  $Re=\num{1000}$ and $\alpha=0.3$ in (c) and (d), respectively. }
		\label{figurefour}
	\end{figure}

	Figure~\ref{figurefour} illustrates the unstable mode of Orr-Sommerfeld
	solution represents by its eigenfunction at critical value of $Re=\num{47145}$ and $\alpha=0.161$. The streamwise velocity fluctuation $u^{\prime}$ reached maximum at $y \sim 0.1576~ \delta$ and then rapidly reduced to 0 at $y \sim 0.1878~\delta$. At the displacement thickness larger than $\delta > 0.18 y$, only small streamwise velocity fluctuation is sustained, less than $7\%$ of the maximum disturbance. For the same mode, we can see that the transverse velocity fluctuation $v^{\prime}$ reached maximum at $y \sim 1.76 ~\delta$. Nevertheless, the amplitude of $v^{\prime}$ will not exceed $10\%$ of the maximum streamwise velocity fluctuation. On the other hand,  at $Re=800$, the maximum $u^{\prime}$ occurred at $y \sim 0.4586~\delta$ and then reduced abruptly to $3\%~\text{of}~ u^{\prime}_{{max}}$ at $y \sim 1.55~\delta$. For the rest of the displacement thickness, the $u^{\prime}$ distribution is almost similar to that in $Re=\num{47145}$. Regarding the $v^{\prime}$ distribution, the maximum $v^{\prime}$ took place at $y\sim 0.1925 ~\delta$ with much larger value than that in $Re=47145$. Beyond this point, the rate of decrease of $v^{\prime}$ is much higher than in the other case. Note that the maximum velocity fluctuations occur very close to the wall compared to the result of the Blasius boundary layer case in Figure \ref{figurefive}.

	Lastly, it is worth mentioning that the critical Reynolds number increased by introducing three-dimensional disturbance in spanwise direction with wavenumber $\beta$. Tabel~\ref{tab1} shows the comparison of the growth rate of the unstable mode between two- and three-dimensional disturbances at $Re=\num{47145}$ and $\alpha=0.161$. For the two-dimensional case, one of the modes has positive $\omega_{i}$, which indicates the temporal amplification of disturbance. In contrast --- in the case of three-dimensional disturbances --- no unstable mode appeared. The results lead to the fact that the flow with two-dimensional disturbance is less stable than that one with three-dimensional disturbance  \cite{reshotko1976boundary}. For instance, at the same $\alpha= 0.161$, the $Re_{{crit}}$ is slightly increase to $\num{47522}$ and $\num{50130}$ for $\beta= 0.02$ and 0.05, respectively. These results showed that a larger spanwise disturbances wavelength promotes the laminar-to-turbulent transition in the asymptotic suction boundary layer.

	\begin{table}[h]
		\centering
		\caption{Comparison of the complex frequency of asymptotic suction boundary layer flow with 2-D and 3-D disturbances.}
		\fontsize{9pt}{9pt}\selectfont
		\begin{tabular}{@{}>{\centering}p{3cm}>{\centering}p{3cm} >{\centering}p{3cm} c @{}}
			\toprule
			\multicolumn{2}{c}{\begin{tabular}[c]{@{}c@{}}Re = \num{47145}, \\ $\alpha=0.161$, $\beta=0$\end{tabular}} & \multicolumn{2}{c}{\begin{tabular}[c]{@{}c@{}}Re = \num{47145}, \\ $\alpha=\num{0.161}$, 	$\beta=\num{0.05}$\end{tabular}}                                                                              \\ \midrule
			$\omega_i$                                                                                                 & $\omega_r$                                                                                                                                       & $\omega_i$                                      & $\omega_r$               \\ \midrule
			\num{-0.009053723135213}                                                                                   & \num{-0.016284398479794}                                                                                                                         & {-0.009053776187042}                            & \num{-0.016284398476526} \\
			~~~\cellcolor{GreenYellow}\num{0.000000016836401}                                                          & \num{-0.025084269698214}                                                                                                                         & \cellcolor{GreenYellow}\num{-0.000040605262884} & \num{-0.025375059688044} \\
			\num{-0.019715168022922}                                                                                   & \num{-0.019808222963637}                                                                                                                         & \num{-0.019829511180454}                        & \num{-0.019898163267561} \\
			\num{-0.015147856968682}                                                                                   & \num{-0.028074047827209}                                                                                                                         & \num{-0.015147910133302}                        & \num{-0.028074047779886} \\
			\num{-0.014162763905649}                                                                                   & \num{-0.039798090915725}                                                                                                                         & \num{-0.019707218199930}                        & \num{-0.037472873998393} \\
			\num{-0.019707164651148}                                                                                   & \num{-0.037472871868033}                                                                                                                         & \num{-0.014074983581941}                        & \num{-0.039941963053997} \\
			\num{-0.025997905784513}                                                                                   & \num{-0.039150965885194}                                                                                                                         & \num{-0.026110370920695}                        & \num{-0.039187669680076} \\
			\num{-0.023416938588254}                                                                                   & \num{-0.045589598709860}                                                                                                                         & \num{-0.023416980778494}                        & \num{-0.045589586407514} \\
			\num{-0.022614038076655}                                                                                   & \num{-0.054023682605554}                                                                                                                         & \num{-0.022543779576827}                        & \num{-0.054118999379730} \\
			\num{-0.026547462476345}                                                                                   & \num{-0.052853339623126}                                                                                                                         & \num{-0.026547479877181}                        & \num{-0.052853108591762} \\
			\num{-0.031150486307261}                                                                                   & \num{-0.054321234640715}                                                                                                                         & \num{-0.031244051799302}                        & \num{-0.054341328451605} \\
			\num{-0.029241864257059}                                                                                   & \num{-0.059487715454088}                                                                                                                         & \num{-0.029242915726632}                        & \num{-0.059487137894433} \\
			\num{-0.028481357069050}                                                                                   & \num{-0.066266959054948}                                                                                                                         & \num{-0.028405253757335}                        & \num{-0.066333628233294} \\
			\num{-0.031588105363357}                                                                                   & \num{-0.065628213232931}                                                                                                                         & \num{-0.031596059809968}                        & \num{-0.065613460818158} \\
			\num{-0.035241231510905}                                                                                   & \num{-0.067069203373296}                                                                                                                         & \num{-0.035320135032351}                        & \num{-0.067076161492354} \\
			\num{-0.033645438888138}                                                                                   & \num{-0.071364920026495}                                                                                                                         & \num{-0.033798694427506}                        & \num{-0.071403462854990} \\
			\num{-0.032867350053255}                                                                                   & \num{-0.077067399963911}                                                                                                                         & \num{-0.032592047402602}                        & \num{-0.075868615970100} \\
			\num{-0.035456352941508}                                                                                   & \num{-0.076761622026041}                                                                                                                         & \num{-0.036988707669012}                        & \num{-0.076440447385795} \\
			\num{-0.038509922447651}                                                                                   & \num{-0.078149971266927}                                                                                                                         & \num{-0.032517962172885}                        & \num{-0.080649864764881} \\
			\num{-0.037052429312565}                                                                                   & \num{-0.081865100218790}                                                                                                                         & \num{-0.039627621987226}                        & \num{-0.078366770248989} \\
			\num{-0.036291429007727}                                                                                   & \num{-0.086355338088544}                                                                                                                         & \num{-0.040517817377491}                        & \num{-0.082570577725276} \\
			\num{-0.038444131642543}                                                                                   & \num{-0.086721482667993}                                                                                                                         & \num{-0.031726915286234}                        & \num{-0.086794104899245} \\
			\num{-0.041097392645718}                                                                                   & \num{-0.087432864575112}                                                                                                                         & \num{-0.040694033753390}                        & \num{-0.087714536935479} \\
			\num{-0.036722939249462}                                                                                   & \num{-0.090703387787520}                                                                                                                         & \num{-0.043935741487887}                        & \num{-0.086387349475554} \\
			\num{-0.047630287666506}                                                                                   & \num{-0.086827461011890}                                                                                                                         & \num{-0.029278753171951}                        & \num{-0.092409107112840} \\
			\num{-0.039887270505808}                                                                                   & \num{-0.091489459185393}                                                                                                                         & \num{-0.036152556100804}                        & \num{-0.090124274796707} \\
			\num{-0.034362792174412}                                                                                   & \num{-0.095332780205598}                                                                                                                         & \num{-0.048148400180181}                        & \num{-0.085015566237441} \\
			\num{-0.040156933113882}                                                                                   & \num{-0.094435258807304}                                                                                                                         & \num{-0.038015640933561}                        & \num{-0.093714379662011} \\
			\num{-0.032332410516983}                                                                                   & \num{-0.100044888343555}                                                                                                                         & \num{-0.050922598385674}                        & \num{-0.087495365371182} \\
			\num{-0.057155897004390}                                                                                   & \num{-0.088413933670037}                                                                                                                         & \num{-0.026715476195033}                        & \num{-0.098064071866723} \\ \bottomrule
		\end{tabular}
		\label{tab1}
	\end{table}

	\section{Conclusion}
	\label{sec:conclusion}

	The temporal stability of the asymptotic suction boundary layer was carried out using the spectral collocation method. For two-dimensional disturbance,  the critical Reynolds number is $Re = \num{47145}$ with streamwise disturbance wavenumber of $\alpha=0.161$. Solving Orr-Sommerfeld equation using spectral collocation gave a good approximation that sufficiently predicts the instability onset. By introducing three-dimensional disturbances, the flow is more stable; hence a slight delay in transition can be achieved. More work is needed to include the porous assumption into the linearized Navier-Stokes equation mathematical model to represent the suction flow better.

	\section*{Nomenclature}
	\begin{tabular}{lcl}
		$a_{n}, T_{n}$  & = & Constants, and polynomial of the Chebychev expansion \\
		$c$             & = & Phase velocity                                       \\
		$N$             & = & Chebychev degree of expansion                        \\
		$Re$            & = & Reynolds number                                      \\
		$U_{o}$         & = & Mean velocity                                        \\
		$u, v$          & = & Streamwise, and transverse velocity disturbances     \\
		$x, y, z$       & = & Streamwise, transverse, and spanwise directions      \\
		$\alpha, \beta$ & = & Streamwise, and spanwise wave numbers                \\
		$\delta$        & = & Displacement thickness                               \\
		$\bar{\eta}$    & = & Nondimensional normal vorticity function             \\
		$\omega$        & = & Disturbance complex frequency                        \\
		$\psi$          & = & Stream function                                      \\
		$\bar{\psi}$    & = & Nondimensional stream function                       \\
		$\varphi$       & = & Freestream disturbance
	\end{tabular}

\bibliographystyle{unsrt}
\bibliography{ressa-temporal-stability.bib}  

\end{document}